\documentstyle[epsfig]{mn}

\newif\ifAMStwofonts

\ifoldfss
  \newcommand{\rmn}[1] {{\rm #1}}

  \ifCUPmtlplainloaded \else
    \NewTextAlphabet{textbfit} {cmbxti10} {}
    \NewTextAlphabet{textbfss} {cmssbx10} {}
    \NewMathAlphabet{mathbfit} {cmbxti10} {} 
    \NewMathAlphabet{mathbfss} {cmssbx10} {} 
  \fi
  \ifAMStwofonts
    \ifCUPmtlplainloaded \else
      \NewSymbolFont{upmath} {eurm10}
      \NewSymbolFont{AMSa} {msam10}
      \NewMathSymbol{\upi}     {0}{upmath}{19}
      \NewMathSymbol{\umu}     {0}{upmath}{16}
      \NewMathSymbol{\upartial}{0}{upmath}{40}
      \NewMathSymbol{\leqslant}{3}{AMSa}{36}
      \NewMathSymbol{\geqslant}{3}{AMSa}{3E}

    \fi
  \fi
\fi 

\ifnfssone
  \newmathalphabet{\mathit}
  \addtoversion{normal}{\mathit}{cmr}{m}{it}
  \addtoversion{bold}{\mathit}{cmr}{bx}{it}
  \newcommand{\rmn}[1] {\mathrm{#1}}

  \newmathalphabet{\mathbfit} 
  \addtoversion{normal}{\mathbfit}{cmr}{bx}{it}
  \addtoversion{bold}{\mathbfit}{cmr}{bx}{it}
  \newmathalphabet{\mathbfss} 
  \addtoversion{normal}{\mathbfss}{cmss}{bx}{n}
  \addtoversion{bold}{\mathbfss}{cmss}{bx}{n}
  \ifAMStwofonts
    \ifCUPmtlplainloaded \else
      \UseAMStwoboldmath
      \makeatletter
      \new@mathgroup\upmath@group
      \define@mathgroup\mv@normal\upmath@group{eur}{m}{n}
      \define@mathgroup\mv@bold\upmath@group{eur}{b}{n}
      \edef\UPM{\hexnumber\upmath@group}
      \new@mathgroup\amsa@group
      \define@mathgroup\mv@normal\amsa@group{msa}{m}{n}
      \define@mathgroup\mv@bold\amsa@group{msa}{m}{n}
      \edef\AMSa{\hexnumber\amsa@group}
      \makeatother
      \mathchardef\upi="0\UPM19
      \mathchardef\umu="0\UPM16
      \mathchardef\upartial="0\UPM40
      \mathchardef\leqslant="3\AMSa36
      \mathchardef\geqslant="3\AMSa3E
    \fi
  \fi
\fi 

\ifnfsstwo
  \newcommand{\rmn}[1] {\mathrm{#1}}

  \DeclareMathAlphabet{\mathbfit}{OT1}{cmr}{bx}{it}
  \SetMathAlphabet\mathbfit{bold}{OT1}{cmr}{bx}{it}
  \DeclareMathAlphabet{\mathbfss}{OT1}{cmss}{bx}{n}
  \SetMathAlphabet\mathbfss{bold}{OT1}{cmss}{bx}{n}
  \ifAMStwofonts
    \ifCUPmtlplainloaded \else
      \DeclareSymbolFont{UPM}{U}{eur}{m}{n}
      \SetSymbolFont{UPM}{bold}{U}{eur}{b}{n}
      \DeclareSymbolFont{AMSa}{U}{msa}{m}{n}
      \DeclareMathSymbol{\upi}{0}{UPM}{"19}
      \DeclareMathSymbol{\umu}{0}{UPM}{"16}
      \DeclareMathSymbol{\upartial}{0}{UPM}{"40}
      \DeclareMathSymbol{\leqslant}{3}{AMSa}{"36}
      \DeclareMathSymbol{\geqslant}{3}{AMSa}{"3E}
    \fi
  \fi
\fi 

\ifCUPmtlplainloaded \else
  \ifAMStwofonts \else 
    \def\upi{\pi}
    \def\umu{\mu}
    \def\upartial{\partial}
  \fi
\fi

\title{The aftermath of the first stars:  massive black holes}
\author[J.L. Johnson and V. Bromm]
       {Jarrett L. Johnson\thanks{E-mail: jljohnson@astro.as.utexas.edu} and Volker Bromm \\
 Department of Astronomy, University of Texas, Austin, TX 78712, USA \\}

\pubyear{2006}

\begin{document}

\maketitle
\topmargin-1cm

\label{firstpage}

\begin{abstract}
We investigate the evolution of the primordial gas surrounding the first massive black holes formed by the collapse of Population III stars at redshifts $z$ $\ga$ 20.  
Carrying out three-dimensional hydrodynamical simulations using GADGET, we study the dynamical, thermal and chemical evolution of the first relic H~II regions.  We also 
carry out simulations of the mergers of relic H~II regions with neighboring neutral minihaloes, which contain high density primordial gas that could accrete onto a 
Pop~III remnant black hole. 
 We find that there may have been a significant time delay, of order $\sim 10^8$~yr, between black hole formation and the onset of efficient accretion. The
build-up of supermassive black holes, believed to power the $z\ga 6$ quasars observed in the {\it Sloan Digital Sky Survey}, therefore faces a crucial 
early bottleneck. More massive seed black holes may thus be required, such
as those formed by the direct collapse of a primordial gas cloud facilitated
by atomic line cooling.
The high optical depth to Lyman-Werner (LW) photons that results from the high fraction of H$_2$ molecules that form in relic H~II regions, combined 
with the continued formation of H$_2$ inside the dynamically expanding relic H~II region, leads to shielding of the molecules inside these regions at least until a 
critical background LW flux of $\sim$ 10$^{-24}$ ergs s$^{-1}$ cm$^{-2}$ Hz$^{-1}$ sr$^{-1}$ is established.  Furthermore, we find that a high fraction of 
HD molecules, $X_{\rmn HD}$ $\ga$ 10$^{-7}$, is formed, potentially enabling the formation of Pop~II.5 stars, with masses of the order of $\sim$ 10 ${\rmn M}_{\odot}$,  during later stages of structure formation when the relic H~II region gas is assembled into a sufficiently deep potential well to gravitationally confine the gas again.
\end{abstract}


\section{Introduction}
What were the feedback effects from the first generation of stars in the Universe?  The first stars, which formed in dark matter (DM) minihaloes of 
mass $\sim 10^{6}{\rmn M}_{\odot}$ at redshifts of $z \sim 20$, were likely very massive, having characteristic masses of the order of $\sim 100 {\rmn M}_{\odot}$ 
(Bromm, Coppi \& Larson 1999, 2002;  Abel, Bryan \& Norman 2002; Nakamura \& Umemura 2001).  These massive Population III stars would have radiated at temperatures of 
$\sim$ 10$^5$~K (e.g. Bond, Arnett \& Carr 1984), generating enough ionizing photons to completely ionize the minihaloes in which they were formed and to contribute to 
the reionization of the Universe (e.g. Barkana \& Loeb 2001; Alvarez, Bromm \& Shapiro 2006).  

In addition to the radiation emitted by the first stars during their lifetimes, 
those Population III stars with masses 40 ${\rmn M}_{\odot}$ $\la$ $M_{*}$ $\la$ 140 ${\rmn M}_{\odot}$ or  $M_{*}$ $\ga$ 260 ${\rmn M}_{\odot}$ are predicted to have 
collapsed to form black holes directly, possibly providing the seeds for the first quasars (Madau \& Rees 2001; Heger et al. 2003; Madau et al. 2004; Ricotti \& 
Ostriker 2004; Kuhlen \& Madau 2005), although 
more massive seed black holes may have been formed after the epoch of the first stars in DM haloes with virial temperatures of $\ga$ 10$^4$ K (Bromm \& Loeb 2003; 
Begelman, 
Volonteri \& Rees 2006; Spaans \& Silk 2006).   

The growth of the first black holes must have been rapid enough to account for the powerful quasars observed at redshifts of $z \ga 6$ (e.g. Fan et al. 2004, 2006), 
believed to be fueled by accretion onto supermassive black holes (SMBHs) with masses $\sim$ 10$^9$ ${\rmn M}_{\odot}$ (e.g. Haiman \& Loeb 2001; Volonteri \& Rees 2005; 
Volonteri \& Rees 2006).  How such vigorous accretion of matter could have taken place poses an important question, as it has been shown that the radiation from the first 
stars heats and evacuates the gas residing within the $\sim$ 10$^6$ ${\rmn M}_{\odot}$ minihaloes in which they are born (Kitayama et al. 2004; Whalen, Abel \& Norman 
2004; Alvarez et al. 2006).  Notwithstanding some possible contribution to the accreted mass from self-interacting dark matter (SIDM) particles (see Spergel \& Steinhardt 
2000; Hu et al. 2006), the baryonic mass around these primordial massive black holes (MBH) must have been efficiently replenished soon after the birth of the black hole.  
In the course of hierarchical structure formation, this continued accretion of matter is naturally accomplished through mergers of the black hole's parent halo with its 
neighboring haloes (see e.g. Ricotti \& Ostriker 2004; Kuhlen \& Madau 2005; Malbon et al. 2006; Li et al. 2006). 
As recent theoretical work on the growth of supermassive black holes has been carried out under the assumption that Pop~III seed black holes can begin accreting at 
the Eddingtion limit very soon after their formation (e.g. Li et al. 2006; Malbon et al. 2006), it stands as an important task to determine in which environments this 
might actually be possible.   

The effects of the radiation from the first stars may have enhanced subsequent star formation through the production of H$_2$ inside relic H~II regions, as well as in 
partially ionized shells just ahead of ionization fronts, as proposed by Ricotti, Gnedin \& Shull (2001, 2002; see also Ahn \& Shapiro 2006).  The former possibility has 
recently received considerable attention (Oh \& Haiman 2003; O'Shea et al. 2005; Nagakura \& Omukai 2005; Johnson \& Bromm 2006). 
O'Shea et al. (2005) have reported that second-generation star formation could have occurred in the ionized minihaloes neighboring the first stars, owing to the formation 
of H$_2$ molecules in the recombining primordial gas (see Shapiro \& Kang 1987; Ferrara 1998; Ricotti, Gnedin \& Shull 2001, 2002).  However, Alvarez et al. (2006) find 
that neighboring minihaloes are self-shielded to the ionizing radiation of the first stars, and thus that star formation in neighboring minihaloes may not, in fact, have 
been significantly enhanced.  Also, it has been shown that the 
activation of cooling by deuterium hydride (HD) molecules inside relic H~II regions may provide an avenue for the formation of Population II.5 (Pop~II.5) stars, with 
masses of the order of 10 ${\rmn M}_{\odot}$ and formed from strongly ionized primordial gas (Mackey, Bromm \& Hernquist 2003; Johnson \& Bromm 2006; see also Nagakura 
\& Omukai 2005).  However, it remains to fully elucidate the formation process of Pop~II.5 stars within the first relic H~II regions if, indeed, the 
radiation from the first stars evacuates the gas contained in their parent haloes and, yet, does not substantially ionize the gas in its neighboring minihaloes. 

Here we present the results of three-dimensional numerical simulations of the recombination of the first relic H~II regions, investigating the possibility of Pop~II.5 
star formation in such regions.  We assume, for this case, the first star to 
have a mass of 100 
${\rmn M}_{\odot}$ and to collapse directly to a black hole.  Additionally, we simulate 
the merger of this parent halo with a neighboring neutral minihalo which has not yet experienced star formation, 
in order to determine the necessary conditions for the black hole to begin accreting gas at the Eddington limit, and so to grow to a mass of 10$^9$ ${\rmn M}_{\odot}$ 
by a redshift of $z$ $\sim$ 6.  
  In future work, we will study the feedback from a pair-instability 
supernova, 
the other possible fate predicted for single primordial stars with masses $\ga$ 100  ${\rmn M}_{\odot}$ (e.g. Rakavy \& Shaviv 1967; Bond, Arnett \& Carr 1984; Heger 
et al. 2003).  
The details of our numerical methodology are given in Section 2.  The 
results of our simulations of the recombination of the relic H~II region appear in Section 3.  Our results from the simulation of the merging of the relic H~II region 
with a pre-collapse halo are presented in Section 4, while the implications of these results for the growth of the remnant black hole appear in Section 5.  
Finally, in Section 6 we summarize our results and present our conclusions.

\section{Methodology}
\subsection{Chemical network}

We employ the parallel version of the GADGET code for our three-dimensional numerical simulations.  This code includes a tree, hierarchical gravity solver combined 
with the smoothed particle hydrodynamics (SPH) method for tracking the evolution of the gas (Springel, Yoshida \& White 2001).  Along with H$_2$, H$_2$$^+$, 
H, H$^-$, H$^+$, e$^-$, He, He$^{+}$, and He$^{++}$, we have included the five deuterium species D, D$^+$, D$^-$, HD and HD$^-$, using the same chemical network as 
in Johnson \& Bromm (2006).

\begin{figure}
\vspace{2pt}
\epsfig{file=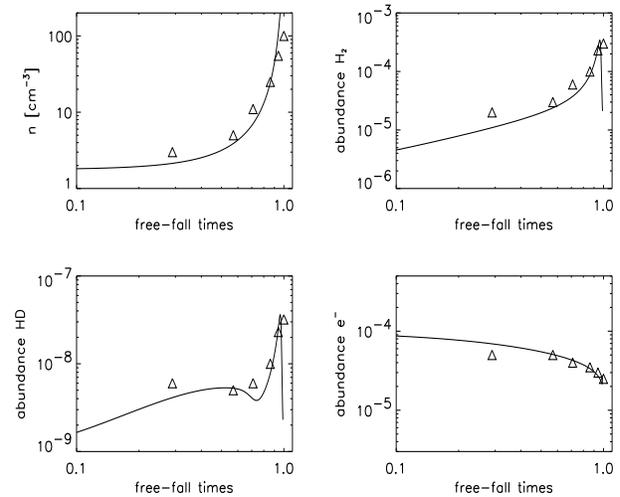,width=8.5cm,height=7.cm}
\caption{Comparison of results from our one-zone model (see Johnson \& Bromm 2006) with the results from a three-dimensional numerical simulation with GADGET, for the 
case of a collapsing spherical minihalo of uniform density.  The solid lines show the results of the one-zone model, while the triangular symbols show output from the 
three-dimensional simulation.  There is clearly good agreement between the two calculations for the evolution of the gas
density (top-left panel), as well as for the H$_2$, HD, and the free electron abundances
 (top-right, bottom-left, and bottom-right panels, respectively). 
}
\end{figure}

As a test of the reliability of the chemical network incorporated into GADGET, we have simulated the idealized case of the homologous collapse of a 
$\sim$ 10$^6$ 
${\rmn M}_{\odot}$ spherical cloud with an initial uniform density of $n_i$ $\sim 2$ cm$^{-3}$. The chemical species are initialized with their primordial abundances, 
as given in Galli \& Palla (1998).  The temperature of the gas, initially a uniform 200 K, was chosen so that there would be little pressure support of the gas against 
gravitational collapse, allowing the cloud to collapse essentially in free-fall.  We then followed the thermal and chemical evolution of the gas near the center of the 
sphere, where the density profile is approximately flat. We compare our simulation results with those obtained 
from the one-zone model, employed in previous work (see Johnson \& Bromm 2006). In this one-zone calculation a cloud 
of uniform density collapses homologously under its own gravity, its density evolving 
according to 

\begin{equation}
\frac{dn}{dt} =  \left(24\pi G \mu m_{\rmn H}\right)^{1/2} n^{3/2}\left[1-\left(\frac{n_{\rmn i}}{n}\right)^{1/3}\right]^{1/2} \mbox{\ ,}
\end{equation}
where $n$ is the density at time $t$ after the onset of the collapse, $m_{\rmn H}$ is the mass of the hydrogen atom, and $\mu$ is the mean molecular weight. 
Fig.~1 shows a comparison of the evolution of the density and of the abundances of H$_2$, HD, 
and free electrons found from our GADGET simulation with that found from the calculation using our one-zone model.  The agreement  
is very good, giving us confidence in the accuracy of our chemical network.

\subsection {First ionizing source}
The initial conditions for our three-dimensional SPH calculation are given by a cosmological simulation of high-$z$ structure formation that evolves both the dark matter 
and baryonic components, initialized according to the $\Lambda$CDM model at $z$ = 100.  In carrying out the cosmological simulation used in this study, we adopt the 
same parameters as in earlier work (Bromm, Yoshida \& Hernquist 2003).  We thus use a periodic box of size $L$ = 100 $h^{-1}$ kpc comoving and a number of particles 
 $N_{\rmn DM}$ = $N_{\rmn SPH}$ = 128$^3$.  The SPH particle mass here is $\sim$ 8 ${\rmn M}_{\odot}$.       

\begin{figure}
\vspace{2pt}
\epsfig{file=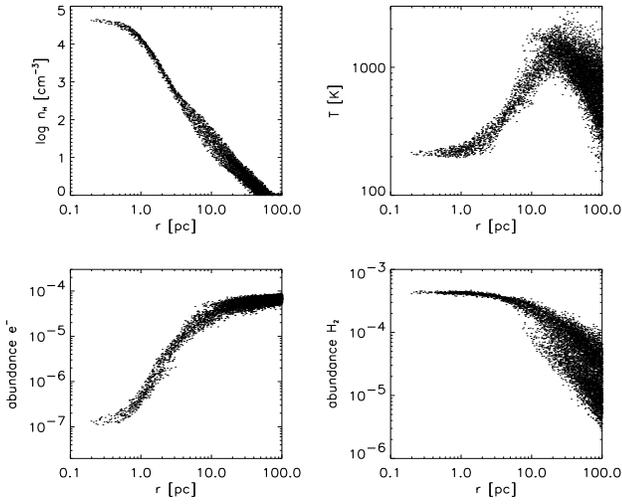,width=8.5cm,height=7.cm}
\caption{The properties of the primordial gas in the minihalo identified to host the first star, as functions of distance from the center. The 
values shown for the central density, temperature, e$^-$ fraction, and H$_2$ fraction are very close to the canonical values generally found in simulations of Pop~III 
star formation (e.g. Bromm \& Larson 2004).        
}
\end{figure}

\begin{figure}
\vspace{2pt}
\epsfig{file=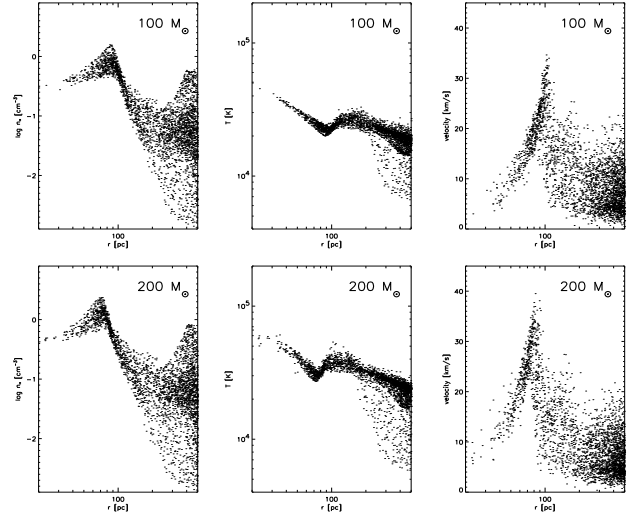,width=8.5cm,height=7.cm}
\caption{The density, temperature and radial velocity of primordial gas ionized and heated by radiation from the first star, as a function of distance from the star, 
for two representative stellar masses.  Here we show the situation after 3 Myr for the case of the 100 
${\rmn M}_{\odot}$ star and after 2 Myr for the case of the 200 ${\rmn M}_{\odot}$ star.  The 100 ${\rmn M}_{\odot}$ star will likely collapse directly to form a 
black hole after 
this time, while the 200 ${\rmn M}_{\odot}$ is likely to explode as a pair-instability supernova.  In the present work, we track the evolution of the ionized gas as 
it recombines and cools 
in the case of the 100 ${\rmn M}_{\odot}$ black hole-forming star.       
}
\end{figure}
We identified the first SPH particle to achieve a density above 10$^{4.5}$ cm$^{-3}$ within our cosmological box at $z \sim 19.5$, finding it at the center of a
minihalo with a total mass of $\sim$ 10$^6$ ${\rmn M}_{\odot}$.  
Fig.~2 presents the properties of the primordial gas as a function of distance from the minihalo center.  The gas temperature rises as particles are 
adiabatically heated as they fall into the potential well of the halo, and then drops nearer the center of the halo where the H$_2$ fraction rises to 
$\sim$ 10$^{-3.4}$ and 
molecular cooling can thus efficiently cool the gas to $\sim$ 200 K (e.g. Bromm \& Larson 2004). 

Having identified the location of the first star, we placed a point source of ionizing radiation at that location in our cosmological box.  This was effected by including 
the following heating rates and ionization rate coefficients in our calculations of the thermal and chemical evolution of the gas:

\begin{equation}
\Gamma_{\rmn HI *} =   n_{\rmn HI} \frac{8.23 \times 10^{-18}} {r^{2}} {\rmn erg\ } {\rmn cm}^{-3} {\rmn s}^{-1}  \mbox{\ }
\end{equation} 

\begin{equation}
\Gamma_{\rmn HeI *} =   n_{\rmn HeI} \frac{1.9 \times 10^{-17}} {r^{2}} {\rmn erg\ } {\rmn cm}^{-3} {\rmn s}^{-1} \mbox{\ }
\end{equation} 

\begin{equation}
\Gamma_{\rmn HeII *} =   n_{\rmn HeII} \frac{3.16 \times 10^{-19}} {r^{2}}  {\rmn erg\ } {\rmn cm}^{-3}  {\rmn s}^{-1} \mbox{\ }
\end{equation} 

\begin{equation}
k_{\rmn H *} =  \frac{8.96 \times 10^{-7}} {r^{2}}  {\rmn s}^{-1}\mbox{\ }
\end{equation} 

\begin{equation}
k_{\rmn HeI *} =  \frac{1.54 \times 10^{-6}} {r^2}  {\rmn s}^{-1}\mbox{\ }
\end{equation} 

\begin{equation}
k_{\rmn HeII *} =   \frac{2.72 \times 10^{-8}} {r^2} {\rmn s}^{-1}\mbox{\ }
\end{equation} 
where $r$ is the distance from the star in pc, and the subscripts denote the chemical species subject to photoionization and photoheating.  
These heating rates and ionization coefficients are derived from the models given in Schaerer (2002) for the case of a $\sim$ 100 ${\rmn M}_{\odot}$ Pop~III star, assuming 
the stars emit a blackbody spectrum (see also Bromm, Kudritzki \& Loeb 2001).  We also carried out a simulation of the ionization and heating of the gas assuming a stellar 
mass of 200 ${\rmn M}_{\odot}$, and the resulting density, temperature, and radial velocity profiles around the central ionizing source, for each case, are shown in 
Fig.~3. The profiles show the situation after 3~Myr and 2~Myr of photoheating and photoionization, for the 100 ${\rmn M}_{\odot}$ and 200 ${\rmn M}_{\odot}$ cases, 
respectively 
(see e.g. Schaerer 2002).  The higher effective temperature and luminosity of the 200 ${\rmn M}_{\odot}$ star results in both a harder spectrum and more ionizing 
photons, and so in a higher heating rate of the surrounding gas from photoionization.  Thus, as can be seen in Fig.~3, the temperature of the gas at a given 
distance from the central source is at least several $\sim$ 10$^3$~K higher for the case of the 200 ${\rmn M}_{\odot}$ star than for that of the 100 
${\rmn M}_{\odot}$ star.  Also, due 
to the shorter lifetime of the 200 ${\rmn M}_{\odot}$ star, the shock that arises from the steep temperature and density gradients encountered during the 
photoheating of the gas has not moved as far out from the central star for the 200 ${\rmn M}_{\odot}$ case, although the shock velocity 
is higher in this case.

Although we neglect the detailed radiative transfer of the ionizing photons here, we succeed at reproducing the basic features of the temperature and density profiles 
of the gas around the ionizing source that have been found in previous radiative transfer calculations (Kitayama et al. 2004; Whalen et al. 2004; Alvarez et 
al. 2006).  Also, we heat and ionize the gas only within 500~pc of the central source, which is roughly consistent with an inhomogeneously ionized region around the 
first star of order $\sim$ 1 kpc, as found by Alvarez et al. (2006), without impinging on neighboring minihaloes in our cosmological box.  We require that our H~II 
region not encompass any neighboring minihaloes because these can be self-shielded to ionizing radiation even if they reside inside the H~II region, and so 
we cannot accurately follow the chemical evolution of the gas inside those minihaloes while the central ionizing source is on.  
Since we do not explicitly follow the propagation of the ionization front with time, we also do not resolve the time-dependent effects on the chemistry and thermal 
evolution of the gas that can give rise to, for instance, the formation of shells of H$_2$ molecules just outside the I-front (see Ricotti, Gnedin \& Shull 2001, 2002).  
Despite the inability of our method to capture the detailed structure of the H~II region, we do expect that we can capture the essential chemical and thermal evolution 
of the relic H~II region as a whole in our simulations, as it expands, cools, and recombines after we remove the central source from the calculation.  

We used the version of 
GADGET which integrates the entropy equation for our photoionization calculation, as we also do later for our simulations of the relic H~II region.  As 
opposed to the integration of the energy equation, this formulation of GADGET conserves both energy and entropy and is much more successful at resolving the 
thermal evolution of gas that experiences shocks or strong local energy injection
 (Springel \& Hernquist 2002).

\subsection {Recombination and molecule formation}
With the formation of a black hole by direct collapse of the 100 ${\rmn M}_{\odot}$ Pop~III star, the relic H~II region left behind begins to recombine and cool.  We 
implement this by simply setting the photoionization coefficients and heating rates to zero, for the case of the 
100 ${\rmn M}_{\odot}$ central star.  Thus, the temperature, density, and radial velocity profiles shown in Fig.~3 are the initial conditions for our simulation of the 
relic H~II region.  We follow the thermal and chemical evolution of the relic H~II region for $\sim$ 100 Myr, considering in particular the production of molecules and the 
cooling of the primordial gas.  

\subsubsection {Photodissociation of molecules}

Molecular hydrogen can easily be dissociated by absorption of Lyman-Werner (LW) photons with energies between 11.2 and 13.6 eV (e.g. Haiman, Rees \& Loeb 1997; Bromm \& 
Larson 2004, and references therein).  Although 
it is well-established that an external LW background could be produced by stars born in neighboring minihaloes
(e.g. Haiman, Rees \& Loeb 1997; Haiman, Abel \& Rees 2000; Ciardi et al. 2000; Machacek, Bryan \& Abel 2001), 
here we assume 
that at the death of our first star a negligible UV background has been established by emission from stars elsewhere in the Universe.  
Thus, to evaluate the effect that photodissociation has on the molecule fraction in the first relic H~II region, we consider as the only 
source of dissociating radiation two-photon emission from the 2$^1$S $\to$ 1$^1$S transition in recombining helium atoms from within the relic H~II region itself 
(Johnson \& Bromm 2006).  Given the much larger Einstein A coefficient for two-photon emission from 2$^1$S than from 2$^3$S, 51.3 s$^{-1}$ for the 2$^1$S $\to$ 1$^1$S 
transition versus 2.2 $\times$ 10$^{-5}$ s$^{-1}$ for the 2$^3$S $\to$ 1$^1$S transition, this should be a sound approximation (Mathis 1957; Osterbrock \& Ferland 2006).

To include a prescription for the photodissociation rates of H$_2$ and HD in our code we assume for simplicity that the relic H~II region is spherically symmetric and 
that it is optically thin to the LW photons.  With this latter assumption, we obtain an upper limit for the dissociation rate, as the molecule fraction can approach 
$\sim$ 10$^{-3}$ in relic primordial H~II regions, which may lead to an appreciable optical depth to LW photons (see e.g. Ricotti, Gnedin \& Shull 2001; 
Oh \& Haiman 2003; Kuhlen \& Madau 2005; 
O'Shea et al. 2005). We estimate the total number of He recombinations, He$^{+}$ + e$^-$ $\to$ He + $h\nu$,
per second within the H~II region that lead to population of the 
2$^1$S state, $Q_{\rmn 2^1 S}$, according to

\begin{equation}
Q_{\rmn 2^1 S} =  \sum \frac{\alpha_{\rmn B} n_{\rmn e} n_{\rmn HeII} m_{\rmn SPH}}{3 \mu m_{\rmn H} n}  \mbox{\ ,}
\end{equation}     
where the sum is over all SPH particles in the H~II region. Here $n_{\rmn e}$ is the number density of free electrons, $n_{\rmn He~II}$ is the number density of He~II, $n$ 
is the total number density, $m_{\rmn SPH}$ is the 
mass per SPH particle, $\mu$ is the mean molecular weight, and $\alpha_{\rmn B}$ is the Case B total He recombination coefficient to singlet states.  We 
take it that $\la$ 1/3 of the recombinations to the singlet levels of He I result 
ultimately in population of the 2$^1$S state, which is accounted for by the factor of 1/3 in the above formula (see Pottasch 1961; Osterbrock \& Ferland 2006). 
For the LW flux at the edge of the H~II region, we find   
 
\begin{equation}
J_{\rmn LW} \sim 10^{-6} \frac{Q_{\rmn 2^1 S}}{4 \pi R^2}  \mbox{\ ,}
\end{equation}  
where $R$ is the radius (in cm) of the He~II region, $J_{\rmn LW}$ is the LW flux in units of 10$^{-21}$ erg s$^{-1}$ cm$^{-2}$ Hz$^{-1}$ sr$^{-1}$, and we have 
conservatively estimated the probability of LW photon emission per two-photon transition 2$^1$S $\to$ 1$^1$S as $\la $ 0.4 (see Osterbrock \& Ferland 2006). We
compute the timescale for the photodissociation of H$_2$ and HD as

\begin{equation}
t_{\rmn diss} \sim 10^8 {\rmn \, yr} \left(\frac{Q_{\rmn 2^1 S}}{10^{45}{\rmn s}^{-1}} \right)^{-1}  \mbox{\ ,}
\end{equation}
where we have taken $R \sim 500$~pc and used $t_{\rm diss}$ $\sim$ 10$^8$ yr ($J_{\rmn LW}$/10$^{-4}$)$^{-1}$ (see Oh \& Haiman 2003; Johnson \& Bromm 2006).  
Inverting this timescale, we find a typical rate for the dissociation of H$_2$ 
and HD of

\begin{equation}
k_{\rmn diss} \sim 10^{-16}{\rmn \,s}^{-1}\left(\frac{Q_{\rmn 2^1 S}}{10^{45}{\rmn s}^{-1}}\right) \mbox{\ ,}
\end{equation}
which is included in the calculation of the molecule fraction in our simulations.

To give an estimate of the importance of the dissociation of molecules due to two-photon emission, we calculate t$_{\rmn diss}$ for the simplified case of a spherical 
recombining H~II region of uniform density. In this case, we have the total number of recombinations to He~I which ultimately result in the 2$^1$S state 
given by 

\begin{eqnarray}
Q_{\rmn 2^1 S} & = & \frac{4 \pi}{9} R^3 \alpha_{\rmn B} n_{\rmn e} n_{\rmn HeII} \nonumber \\
 & \simeq  & 
10^{50}{\rmn \,s}^{-1}\left(\frac{R}{500{\rmn \, pc}}\right)^3
\left(\frac{n_{\rmn e} n_{\rmn HeII}}{{\rmn cm}^{-6}}\right)
\mbox{\ .}
\end{eqnarray}
Using equation (10), we find
\begin{equation}
t_{\rmn diss} \sim 
10^{3}{\rmn \,yr}
\left(\frac{n_{\rmn e} n_{\rmn HeII}}{{\rmn cm}^{-6}}\right)^{-1}
\mbox{\ ,}
\end{equation}
which clearly shows that at high densities and at times when He~recombination is still ongoing, the photodissociation of any molecules that 
have formed may be important, even in the absence of an externally generated LW background.  However, for our case of a relic H~II region surrounding a minihalo, 
since it is predominantly at lower temperatures ($\la$ 5,000 K) that molecules are formed, at which times much of the He II has already recombined, and since our H~II 
region is expanding to lower densities with time, we expect that dissociation of molecules due to two-photon emission will be unimportant, at least 
during the later evolutionary stages.

Molecular hydrogen could also be dissociated by radiation generated during the accretion of gas onto the remnant black hole.  However, as we show in Section 5, we find 
that the accretion rate onto the black hole is low, comparable to that found by O'Shea et al. (2005), for at least a few
10~Myr after the collapse of the central ionizing star.  O'Shea et al. estimate
that this accretion rate results in a photodissociation rate of H$_2$ which is at least an order of magnitude below the formation rate of H$_2$ in the relic H~II 
region.  We thus neglect the possible effects of photodissociating radiation due to accretion onto the remnant MBH.

\subsection {Merging minihaloes}
In order to determine which, and how quickly, neighboring minihaloes will collapse following the formation of a Pop~III remnant black hole, we consider the situation 
in which the relic H~II region surrounding the remnant black hole merges with a neighboring DM halo and its accompanying, un-ionized and dense gas component.  Since 
we are simulating the evolution of the relic H~II region and the infalling neutral minihalo after the death of the Pop~III star, in these merger simulations we do not include 
any photoheating or photodissociating terms in our calculations of the thermal and chemical history of the gas. We initiate 
the merger by placing the spherical relic H~II region, immediately following the collapse of the central star, with properties shown in Fig.~3, adjacent to a second spherical 
region of radius 500~pc at the center of which is a minihalo which is still neutral, not yet having hosted star formation.  
The region containing the neutral minihalo is selected and cut out from elsewhere in the same cosmological box,
and is then placed adjacent to the relic H~II region in a new, otherwise empty simulation box.  
The initial separation between the centers of the two minihalos is 1~kpc proper
for all merger simulations carried out here.

The reason why we must choose these initial conditions, and why we may not simply continue running our cosmological simulation of the relic H~II region and wait for 
a merger to occur, is that we are limited by the size of the cosmological box. Our box size is $\sim$ 100$h^{-1}$~kpc, which is 
too small to contain the large wavelength density modes that drive the mergers of minihaloes. Thus, we carry out the mergers in an empty box, imparting a 
relative velocity of 8~km~s$^{-1}$ to the merging halos, comparable to the virial velocity of the resulting system,
setting them on trajectories for a direct collision.  

We assume that the molecules in the infalling, 
pre-collapse halo have been destroyed by the H$_2$ photodissociating LW flux from 
the nearby Pop~III star which has formed the relic H~II region.  We carry out simulations in which the neighboring halo has peak gas densities of $\sim$ 0.1 
cm$^{-3}$, 1 cm$^{-3}$, 10 cm$^{-3}$, and 100 cm$^{-3}$, corresponding to different degrees of pre-collapse. In the case of the 0.1 cm$^{-3}$ peak density neighboring halo, the free-fall time will be comparable to the 
timescale for the completion of the merger and so we can neglect the possibility of this 
minihalo collapsing to form a star before the completion of the merger.  However, given that we assume that there are no molecules in the neighboring halo at the 
outset of the merger, we expect that collapse will be delayed even for the higher density cases, as cooling will be suppressed until molecules have reformed (see 
Mesinger et al. 2006).  

 Although the initial conditions for these simulations are idealized, as we have not followed the merger of the haloes in a fully cosmological context but instead in a 
box containing only the relic H~II region and the infalling minihalo, we are able to discern the crucial aspects in the thermal and chemical evolution of the gas in 
the vicinity of the remnant Pop~III black hole.
We do note, however, that additional effects that we do not consider here, such as the formation of an H$_2$ shell and the driving of a shock through a partially 
ionized minihalo, can become important for cases in which significant portions of the infalling minihalo are ionized (see Ahn \& Shapiro 2006). A more sophisticated 
treatment of radiative transfer will be required to more 
accurately follow the evolution of the gas
(see Ahn \& Shapiro 2006; Susa \& Umemura 2006).

\section {Evolution of primordial gas in relic H~II regions}
The results of our recombination simulation are presented in Fig.~4 at three representative times after the death of the central star. As can be seen in the panels 
showing the temperature as a function of the density, the gas initially cools largely by adiabatic expansion, as at temperatures below 
$\sim$ 10$^4$ K the gas closely follows the adiabatic relation $T$ $\propto$ $n^{2/3}$, delineated in the panels on the right-hand side.  As the gas recombines, the 
cooling rate due to collisional excitation of the newly-formed hydrogen atoms is enhanced and the temperature of the gas drops to $\sim$ 10$^4$ K, at which point the 
cooling rate decreases when molecular hydrogen becomes the main coolant, aside from adiabatic cooling, which continues as the gas expands into the intergalactic
medium (IGM).  

\begin{figure*}
\includegraphics[width=7.in]{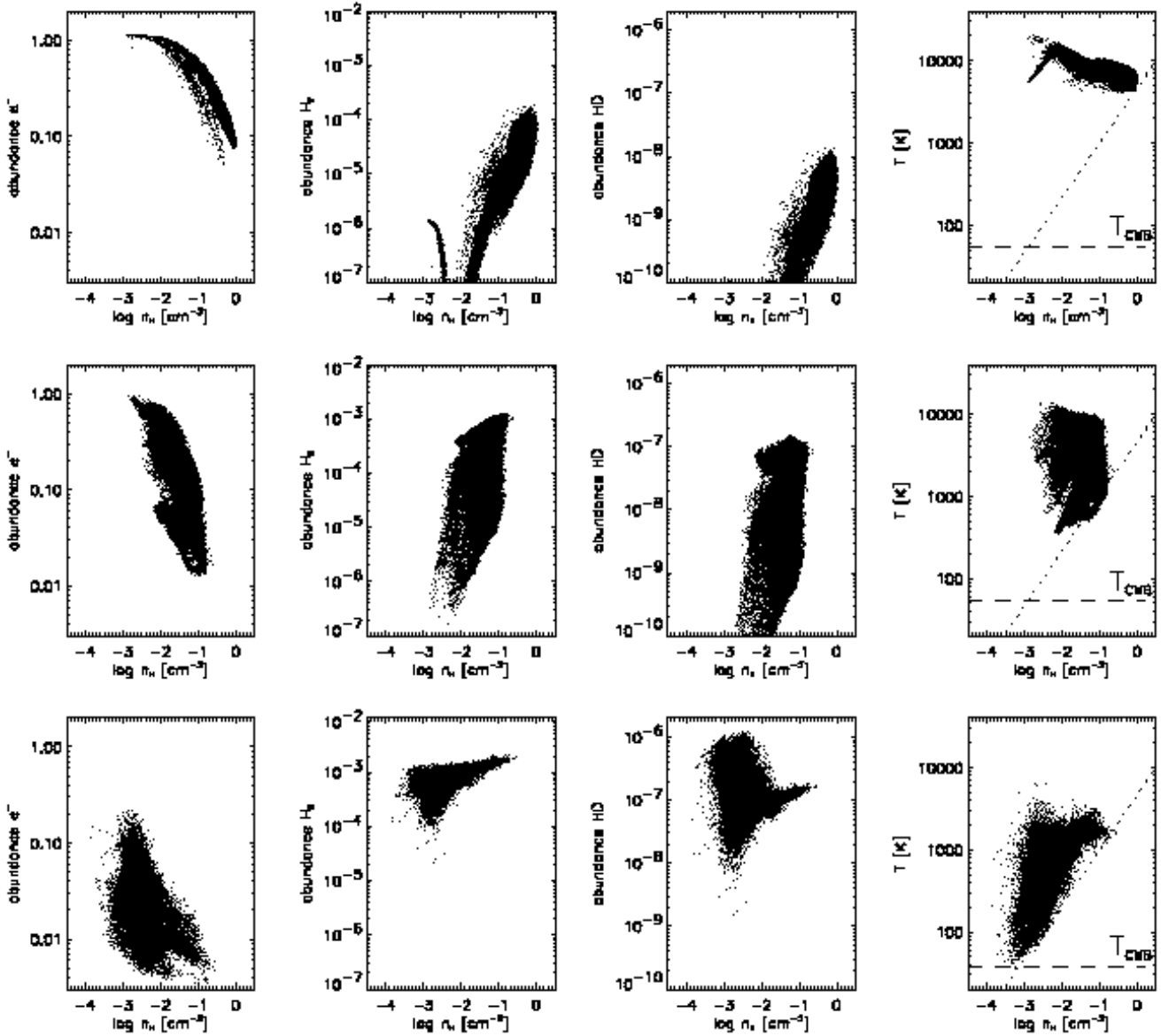}
\caption{The evolution of the relic H~II region.  From left to right, the panels show the free electron fraction, the H$_2$ fraction, the HD fraction, and the 
temperature as functions of density at $\sim$ 1 Myr (top row), $\sim$ 10 Myr (middle row), and $\sim$ 100 Myr (bottom row) after the collapse of the central star to a 
black hole.  The long-dashed line in the rightmost panels denotes the temperature of the cosmic  background radiation, $T_{\rmn CMB}$, while the short-dashed line denotes 
the line $T$ $\propto$ $n^{2/3}$, along which gas evolves adiabatically.  Here, we plot only the SPH particles that were subjected to the photoionizing radiation of the 
central star, that is, particles within $\sim$ 500 pc of the central star.  
}
\end{figure*}
The HD fraction in the highest density regions of the recombining gas increases to $X_{\rmn HD}$ $\sim$ 10$^{-7}$ after $\sim$ 10 Myr since the death of the central star, 
and to 
$X_{\rmn HD}$ $\sim$ 10$^{-6}$ after 100 Myr.  Thus, the HD fraction quickly rises above the critical value of $X_{\rmn HD, crit}$ $\sim$ 10$^{-8}$ for efficient cooling 
of primordial gas in local thermodynamic equilibrium (LTE) to the temperature of the cosmic microwave background (CMB) (see Johnson \& Bromm 2006). Because LTE can only 
be achieved at much higher densities than those that persist in the relic H~II region we consider here, radiative cooling to the CMB floor would only 
be a viable possibility if densities are somehow increased to the point that LTE can be established.  This would 
happen if the gas is at some point incorporated into a larger DM halo and becomes gravitationally bound once more.  That the HD fraction is so high, however, 
means that the potential for Pop~II.5 star formation does exist, in principle, if the gas becomes gravitationally bound and thus available for star formation (see also 
Nagakura \& Omukai 2005).     

We evaluate the recombination time after 100~Myr by taking the density of H$^+$ and of free electrons to be $n_{\rmn H^+}$ $\sim$ $n_{\rmn e} 
\sim 
10^{-4}$cm$^{-3}$, as can be seen from the lower-left panel of Fig.~4. Then, assuming at these low densities a Case A 
recombination coefficient of 
$\alpha_A$ $\sim$ 6 $\times$ 10$^{-13}$ cm$^{3}$ s$^{-1}$, we obtain a recombination time of $t_{\rmn rec} \sim$ 500 Myr.  This is more than twice the Hubble time at 
these redshifts, and suggests that the free electron fraction left over from the ionization caused by the first stars may have remained an important catalyst for molecule 
formation even after hundreds of millions of years since the death of the central star (see e.g. Shapiro \& Kang 1987; Yamada \& Nishi 1998; 
O'Shea et al. 2005; Nagakura \& Omukai 2005).

The optical depth to LW photons becomes unity for H$_2$ column densities of $\sim$ 10$^{14}$ cm$^{-2}$ (e.g. Draine \& Bertoldi 1996; Osterbrock 
\& Ferland 2006), and for our 
relic H~II region we estimate

\begin{equation}
\tau_{\rmn LW} \sim \frac{n_{\rmn H} X_{\rmn H_{2}} R}{10^{14} {\rmn cm}^{-2}}  \mbox{\ ,}
\end{equation} 
where $X_{\rmn H_2}$ is the molecule fraction, $R$ is the radius of the relic H~II region, and $n_{\rmn H}$ is the number density of hydrogen nuclei.  We find that the 
molecule fraction approaches $X_{\rmn H_2}$ $\sim$ 10$^{-3}$ and that the number density becomes $n_{\rmn H}$ $\sim$ 10$^{-3}$cm$^{-3}$.
Taking the radius of the H~II region to be $\sim$ 1 kpc, we find that the optical depth to LW photons becomes of the order of $\tau_{\rmn LW}$ $\sim$ 10.  If the 
density of 
star-forming minihaloes is thus high enough, an appreciable suppression of the background LW flux may result from the high molecule fraction which arises in relic 
H~II regions.  This may provide an important degree of shielding from molecule-dissociating radiation and may lead to a higher overall efficiency of star-formation 
in minihaloes with virial temperatures $\la$ 10$^4$ K (see Ricotti, Gnedin \& Shull 2001; Machacek, Bryan \& Abel 2001, 2003; Oh \& Haiman 2002).

We carried out simulations both with and without the photodissociating two-photon emission from recombining He II included, 
and we found that this photodissociating radiation had little effect on the molecule abundances.  
To estimate the level of LW background radiation necessary to efficiently
photodissociate H$_2$ molecules, we evaluate the H$_2$ formation timescale at $\sim$ 10 Myr after the death of the central star.  Taking representative 
values for the temperature, number density, and abundances of the chemical species after 10 Myr of recombination, we find a formation timescale for H$_2$ of 
$t_{\rmn form}$ $\sim$ 10$^7$ yr.  The continued formation of H$_2$ is driven by the high abundances of H, H$^{-}$, and e$^{-}$, which are the 
reactants in the following 
reaction sequence that is the most important for the production of H$_2$ (e.g. Kang \& Shapiro 1992):

\begin{equation}
{\rmn  H} + {\rmn e^{-}} \to {\rmn H^{-}} + h\nu \mbox{\ ,}
\end{equation} 
       
\begin{equation}
{\rmn  H} + {\rmn H^{-}} \to {\rmn H_{2}} + {\rmn e^{-}} \mbox{\ .}
\end{equation}
Equating the formation timescale for H$_2$ with the dissociation timescale for H$_2$, given by $t_{\rmn diss}$ $\sim$ 10$^8$ yr ($J_{\rmn LW}$/10$^{-4}$)$^{-1}$, 
we find a critical value for the 
background LW flux, below which molecules in the relic H~II region will not be photodissociated efficiently, of the order of $J_{\rmn LW, crit}$ $\sim$ 
10$^{-3}$ (see also Oh \& Haiman 2003).  This is comparable to the background LW flux that is expected to have been established by the first generations of 
stars at redshifts $z \ga 15$ 
(e.g. Greif \& Bromm 2006).  Taking into account the optical depth to LW photons of order $\tau_{\rmn LW}$ $\sim$ 10, we find that the most heavily self-shielded 
molecules in the center of relic H~II region could only be dissociated by a background LW flux, emanating from outside the relic H~II region, at least of the order of 
$J_{\rmn LW}$ $\sim$ 10.  This value would, however, decrease with time if the molecules nearer to the periphery of the relic H~II region were dissociated 
by the external LW background flux, and so become unavailable for shielding the inner molecules from the dissociating radiation. 
We note also that if the gas in the relic H~II region has large velocity gradients, owing to turbulence that we do not resolve in these simulations, then the optical 
depth to LW photons may be lower than the value we find here by a factor of a few (see Draine \& Bertoldi 1996; Osterbrock \& Ferland 2006).  This would not, however, 
effect the value that we find for $J_{\rmn LW, crit}$, as this is independent of the optical depth to LW photons.

The values that we find for $J_{\rmn LW, crit}$ and for $\tau_{\rmn LW}$ suggest that the enhanced fractions of H$_2$ and HD inside relic H~II regions could have 
persisted down to at least redshifts of $z$ $\sim$ 15, and so, in 
principle, would have been available for star formation at least down to these redshifts 
(see also Ricotti, Gnedin \& Shull 2002).  
The high HD abundance, which becomes at least an order of magnitude 
above the critical abundance for cooling the primordial gas to the CMB temperature floor, could thus have led to Pop~II.5 star formation inside the first relic H~II 
regions, if these regions became incorporated into more massive DM haloes that could gravitationally bind, and so increase the density of, the recombining 
primordial gas.     

\section{Evolution of primordial gas in merging minihaloes}
To discern the conditions under which the Pop~III remnant black hole could efficiently accrete the dense, cold gas supplied by a neighboring minihalo, 
we have tracked the evolution of the gas in such merging systems with a range of initial peak densities 
of the gas within the infalling neutral minihalo. 
Again, here we have assumed that the LW flux from the now-collapsed Pop~III star has destroyed all of the molecules inside the neutral infalling halo, although 
molecules can reform during the merger, since there is no longer a LW flux from the original Pop~III star.  
The initial peak 
densities of the halos we follow 
are 0.1, 1, 10, and 10$^2$ cm$^{-3}$.  Fig.~5 shows the evolution of the density structure in the merger between the relic H~II region and a halo with a peak 
density of 10~cm$^{-3}$, as a representative case. 

Fig.~6 shows the time evolution of a merger 
of a relic H~II region, in which the central star has just collapsed to form a black hole, with a neutral pre-collapse minihalo which has a 
peak density of 0.1~cm$^{-3}$ at the time of the formation of the black hole,
at four representative times.  As in 
the case of the recombination of the relic H~II region evolved in our $100 h^{-1}$~kpc cosmological box, which did not experience a merger, the relic H~II region gas 
expands and cools largely adiabatically.
This expansion is evident in Fig.~5 as well.  Also, as can be seen from the temperature rise in the pre-collapse halo, shown in black in Fig.~6, the expansion of the 
relic H~II region, combined with the 8~km~s$^{-1}$ relative velocity of the infalling halo with respect to the relic H~II region, shock-heats the neutral gas 
within the infalling halo, contributing 
to the suppression of the density in the pre-collapse halo.  The gas in the pre-collapse halo, furthermore, does not reform H$_2$ molecules efficiently, 
owing to the low density of the gas in this halo.  The H$_2$ fraction approaches only $\sim$ 10$^{-6}$ after 100 Myr since the death of the central star.
Thus cooling of the gas is inefficient, and the 
highest densities achieved in this merger are $n \la 10^{-0.5}$~cm$^{-3}$.  
\begin{figure*}
\includegraphics[width=7.in]{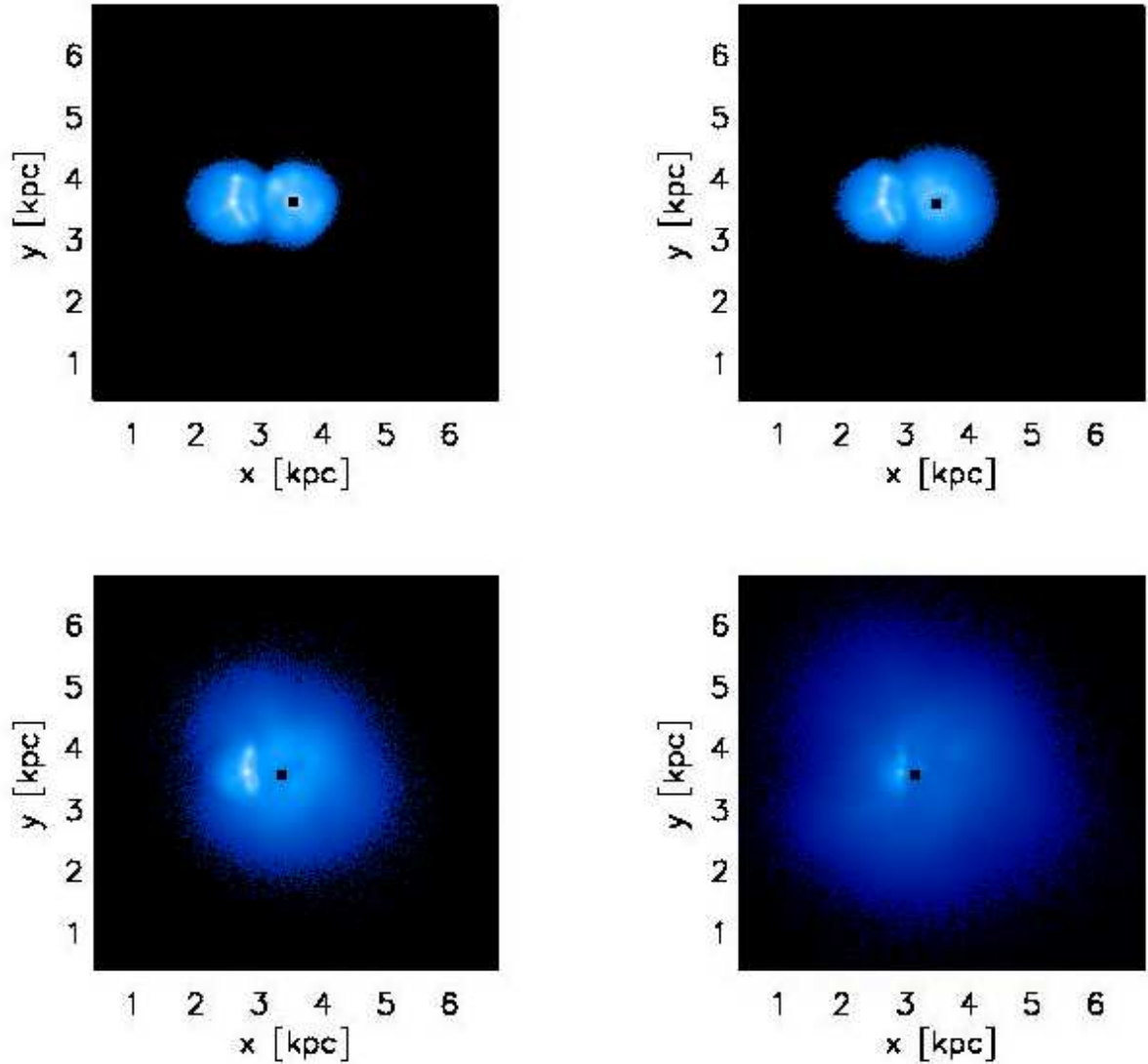}
\caption{
The merging of the relic H~II region with a neutral neighboring minihalo with an initial central density of $\sim$ 10 cm$^{-3}$.  The H~II region is on the right-hand 
side of the top-left panel at the beginning of the merger, while the pre-collapse halo with which it merges is on the left-hand side.  The location of the remnant black 
hole, initially at the center of the relic H~II region, is shown by the black square in each panel.  Here, we assume that the black hole has a ballistic trajectory, 
with a constant velocity of 4~km~s$^{-1}$ to the left, equal to the initial velocity of the relic H~II region.    The highest density gas is shown in white and the lower 
density gas is shown in blue. 
The merger is shown at 1 Myr (top-left), 10 Myr (top-right), 50 Myr (bottom-left), and 100 Myr (bottom-right) after the death of the central star.
The halo shown on the left-hand side collapses to a density of $\sim 10^3$~cm$^{-3}$ during the merger (see Fig.~7).
}
\end{figure*}

We can expect the gas within a minihalo that merges with a relic H~II region to be dispersed and have its density suppressed whenever the ram pressure of the expanding relic H~II 
region, given by $P_{\rmn ram} \sim n_{\rmn H~II} m_{\rmn H} v^2$, is 
higher than the pressure of the neutral gas within the infalling halo, given by $P_{\rmn gas} = n_{\rmn gas} k_{\rmn B} T$, where 
$v$ is the expansion velocity of the relic H~II region, $n_{\rmn H~II}$ is the density in the relic H~II region, $n_{\rmn gas}$ is the density of the gas 
within the neutral minihalo that is 
merging with the relic H~II region, and $T$ is the temperature of this gas.  Taking a fiducial value of $v$ $\sim$ 10~km~s$^{-1}$ for the expansion velocity, as can 
be seen in Fig.~3, and taking $n_{\rmn H~II}$ $\sim$ 10$^{-2}$ cm$^{-3}$ as a typical value for the density of the relic H~II region, 
as can be seen in Fig.~4, we find the 
following condition for the retention of the neutral gas in the potential well of the infalling halo:
\begin{equation}
n_{\rmn gas} T \ga 10^2 {\rmn K\,} {\rmn cm}^{-3} \mbox{\ .}                 
\end{equation} 
The condition in equation (17) can be satisfied for
primordial gas collapsing in a minihalo with a mass of $\sim$ 10$^6$ ${\rmn M}_{\odot}$, provided that  
densities of at least $n$ $\sim$ 10$^{-0.5}$~cm$^{-3}$ have been reached prior
to the merger [see Fig.~10 in Bromm et al. (2002)].

However, if the gas in the infalling minihalo collapses to form a star before the completion of the merger, then the gas in the minihalo will be heated and expand to lower 
densities, as shown in Fig.~3 (see also Abel, Wise \& Bryan 2006).  This final collapse to form a star will occur on the order of the free-fall time, 
$t_{\rmn ff}$ $\propto$ $n^{-1/2}$, if the free-fall time is longer 
than the cooling time (e.g. Tegmark et al. 1997; Ciardi \& Ferrara 2005).  If the gas cannot cool efficiently, however, this collapse will be delayed.  This may 
indeed be the case in regions near the first stars where 
the LW 
flux generated during their $\la$ 3 Myr lives could destroy the H$_2$ molecules inside the pre-collapse haloes, depriving them of the coolants that allow for star 
formation (e.g. Yoshida et al. 2003). 

\begin{figure}
\vspace{2pt}
\epsfig{file=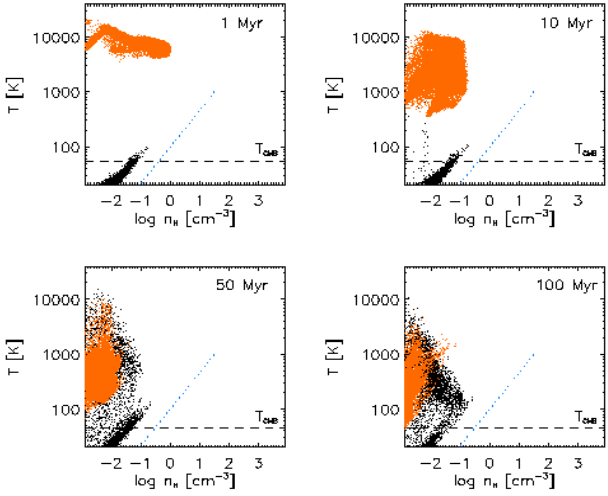,width=8.5cm,height=7.cm}
\caption{
The evolution of the merger of the relic H~II region with a neutral neighboring minihalo.  The H~II region gas is in orange, while the gas from the pre-collapse minihalo 
is in black.  The central density of the pre-collapse minihalo is initially $\sim$ 0.1 cm$^{-3}$, roughly 10$^2$ times the background IGM density.  We follow the same 
convention for the lines
 delineating the CMB temperature and adiabatic evolution as in Fig.~4.
The criterion given by equation (17) for neutral gas retention is not 
satisfied in this case, 
and the low density gas in the infalling halo is shock-heated and remains at low densities during the merger.  Furthermore, the H$_2$ fraction stays below 10$^{-6}$ for 
the entire 100 Myr duration of the merger, further preventing the gas from cooling and collapsing to higher densities. 
}
\end{figure}

\begin{figure}
\vspace{2pt}
\epsfig{file=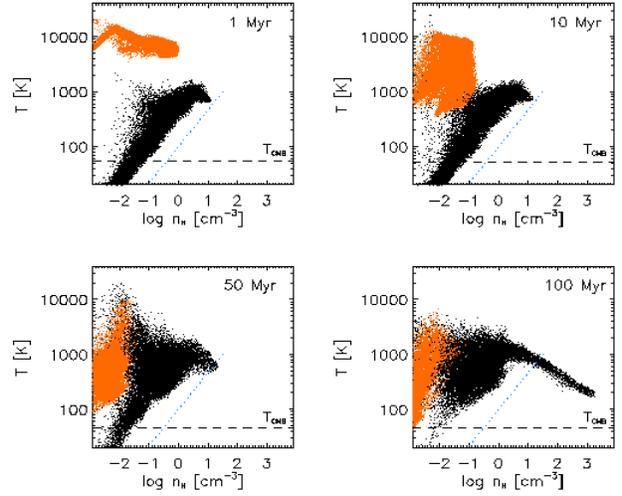,width=8.5cm,height=7.cm}
\caption{
Same as in Fig.~6, except that the neutral gas has an initial peak density of $\sim$ 10 cm$^{-3}$.  The 
criterion in equation (17) for neutral gas retention is satisfied here, 
and the densest gas, in the center of the infalling minihalo, retains its high density despite the outer layers of gas being shock-heated in the merger.  
Also, the H$_2$ fraction approaches 10$^{-3}$ after 
100 Myr at these higher densities, so that the gas in the merging halo cools and collapses efficiently to a density of $\sim$ 10$^3$ cm$^{-3}$.  As the labels 
indicate, the four panels correspond to the same times since the death of the central star as the four panels in Fig.~5, which shows the evolution of the projected 
gas density. 
}
\end{figure}

To find the highest densities that could be achieved during a merger of a neutral minihalo with a relic H~II region, we compare the timescale for completion of the merger, 
$t_{\rmn merge}$, with the timescale for the collapse of the neutral minihalo, $t_{\rmn collapse}$, the latter found from our simulations of mergers involving minihaloes 
with peak densities of 1, 10, and 10$^2$ cm$^{-3}$ at the time of the formation of the black hole and the cessation of the radiation from the original Pop~III star.  
For each of these initial densities, the criterion given by equation (17) for neutral gas retention is satisfied.
Cases for which $t_{\rmn merge} \la t_{\rmn collapse}$ will give rise to mergers resulting in the highest densities of gas that can be accreted onto the black hole, 
as it is these minihalos that will merge completely with the black hole before collapsing to form a second star. We define the merger timescale as

\begin{equation}
t_{\rmn merge} \simeq \int_{r_{\rmn MBH}}^0      \frac{dr}{(2{\rmn G} M_{\rmn halo} (\frac{1}{r}-\frac{1}{r_{\rmn MBH}}))^{\frac{1}{2}}} \mbox{\ ,}                 
\end{equation}
where $r_{\rmn MBH}$ 
is the initial distance between the centers of the merging haloes at the time of the formation of the massive black hole in the relic H~II region, and $M_{\rmn halo}\sim 10^6 {\rmn M}_{\odot}$. 
We note, however, that this is only an approximation to the actual time that would be required for a merger to take place, 
as this formula assumes that the merging minihaloes start at rest with respect to each other at the time of the collapse of the first star, and this will not be the 
case in general (see e.g. Abel, Wise \& Bryan 2006). 

Fig.~7 shows the time evolution for a merger with a neutral peak density of 10 cm$^{-3}$, at four 
representative times, just as in Fig.~6.  The gas in the pre-collapse halo, in this case, does reform H$_2$ molecules efficiently, owing to the high density of the gas 
in this halo.  The H$_2$ fraction approaches $\sim$ 10$^{-3}$ after 100 Myr. Thus, gas cooling is efficient, and the density reaches 
$n$ $\sim$ 10$^{3}$ cm$^{-3}$ after $\sim$ 100 Myr. Fig.~5 shows the evolution of the gas density structure in this merger, at the 
four representative times which are also shown in Fig.~7.

For the case of a merger with an initial peak density of 1 cm$^{-3}$, we find that the H$_2$ fraction becomes only of the order of 
10$^{-5}$, and the density in this minihalo thus does not increase beyond $\sim$ 1 cm$^{-3}$ within 100 Myr, since molecular cooling is suppressed.  We also find 
that for the case of a merging minihalo with an initial density of 100 cm$^{-3}$ the H$_2$ fraction becomes $\sim$ 10$^{-3}$ and the halo collapses after $\sim$ 60 Myr.  
As expected, for higher initial peak densities in the infalling haloes, the timescales for the collapse of these haloes become shorter, both because H$_2$ molecules are 
reformed more quickly and because the free-fall time is shorter for denser haloes (see also Mesinger et al. 2006).

\begin{figure}
\vspace{2pt}
\epsfig{file=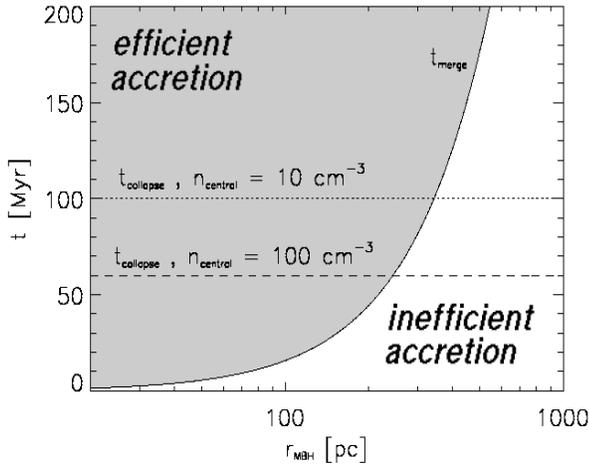,width=8.5cm,height=7.cm}
\caption{Requirements for the efficient accretion of gas onto a Pop~III remnant black hole. The two horizontal lines show the time it takes for 
infalling minihaloes with central densities of 10 and 100~cm$^{-3}$ to collapse and form stars.  The merger timescale, $t_{\rmn merge}$, defined in equation 
(18), is the time it takes for the remnant black hole to merge with the infalling minihalo, and is a function of $r_{\rmn MBH}$, the distance between the black hole 
and the center of the infalling minihalo at the time of the formation of the black hole.  For the black hole to efficiently accrete gas at near the Eddington rate, 
it must merge with the infalling halo before this halo collapses to form a star.  Thus, efficient accretion onto the black hole is only possible if 
$t_{\rmn merge} <
t_{\rmn collapse}$.
}
\end{figure}

Fig.~8 shows the requirements for infalling minihaloes to collapse to high densities on a timescale that is longer than the timescale for their merging 
with a black hole inside a relic H~II region.  These infalling minihaloes, which have $t_{\rmn merge} \la t_{\rmn collapse}$, are those that will contain high 
density gas ($\ga 10^2$~cm$^{-3}$) at the completion of the merger, which can lead to efficient accretion onto the black hole.  
As can be seen in Fig.~8, at the time of the death of the original Pop~III 
star, these minihaloes have densities $\ga 10$~cm$^{-3}$ and lie within $r_{\rmn MBH}$ $\sim 340$~pc of the halo hosting the first 
Pop~III star. Higher densities than this could be achieved, within a merger timescale of the order of $\sim$ 100 Myr and without forming a second star, in mergers with 
minihaloes that have initial densities of the order of 10$^2$ cm$^{-3}$ only 
if the timescale for completion of the merger is $t_{\rmn merge}$ $\la$ 60 Myr, the time in which a 10$^2$ cm$^{-3}$ peak density halo would collapse to form a star.  
As Fig.~8 shows, this would demand that such a dense minihalo be within $r_{\rmn MBH}$ $\sim$ 240 pc of the halo hosting the first Pop~III star at the time that this 
first Pop~III 
star collapses to form a black hole.  Such minihaloes would lie well within the H~II region of the Pop~III star and so may be strongly heated and partially ionized.  
However, the gas in 
such a halo is expected to remain neutral at densities $n$ $\ga$ 2 cm$^{-3}$ (Alvarez et al. 2006; see also Abel, Wise \& Bryan 2006; Susa \& Umemura 2006), 
which justifies the assumption that we use in our simulations of neutral neighboring minihaloes 
adjacent to the minihalo hosting the first Pop~III star, for haloes containing gas at densities $\ga$ 
10 cm$^{-3}$.

\section{Black hole accretion and feedback}
In order for a black hole with a mass of $\sim 100 {\rmn M}_{\odot}$, such as may have been formed from the direct collapse of a Pop~III star at $z$ $\sim$ 
20 (e.g. Heger et al. 2003), to attain a mass of $\sim$ 10$^{9}$ ${\rmn M}_{\odot}$, such as have been inferred for the black holes that power quasars at $z$ $\ga$ 6, 
an average accretion 
rate of $\ga$ 1 ${\rmn M}_{\odot}$ yr$^{-1}$ is required.  Here we analytically estimate the accretion rate of a black hole that begins with a mass $m_{\rmn i}$, the resulting 
mass of the black hole $m$ as a function of time $t$ since its formation, and the luminosity that would be produced in the process of this accretion as a function 
of time, for a given density and temperature of accreted gas.  We assume that the black hole accretes gas from a cloud with uniform density and temperature and 
with dimensions much greater than the accretion radius of the black hole, $r_{\rmn acc}$ $\sim$ G$m$/$c_{\rmn s}^{2}$, where $c_{\rmn s}$ is the speed of sound in 
the gas.  In this case, we can estimate the Bondi-accretion rate as (Bondi 1952)           

\begin{equation}
\dot{m}_{\rmn B} \sim \frac{2 \pi(Gm)^2 m_{\rmn H} n}{c_{\rmn s}^3}\mbox{\ .}
\end{equation} 
Assuming that a fraction $\epsilon\simeq 0.1$ of the accreted mass is converted into energy and radiated away, so that the black hole grows according to 
$dm/dt\simeq (1-\epsilon)\dot{m}_{\rmn B}$, integrating from $t$=0 and $m=m_{\rmn i}$ yields a time dependent 
mass of the black hole of

\begin{equation}
m \sim \left(\frac{1}{m_{\rmn i}} - \frac{2 \pi (1-\epsilon)m_{\rmn H}^{\frac{5}{2}}   G^2  n t}{(3 k T)^{\frac{3}{2}}   }\right)^{-1}  \mbox{\ ,}
\end{equation}
where $T$ is the temperature of the gas at infinity and where we have used $c_{\rmn s}^2 = 3 k_{\rmn B} T$/$m_{\rmn H}$.  

Using equation (19) for the mass accretion rate, we compute the luminosity $L$ generated in the process of the black hole accreting mass, operating as a miniquasar, as

\begin{equation}
L = \frac{\epsilon}{1 - \epsilon} \frac{dm}{dt}c^2 = \frac{2  \pi \epsilon (Gm)^2 m_{\rmn H} n c^2}{(1 - \epsilon)c_{\rmn s}^3}\mbox{\ ,}
\end{equation} 
where $c$ is the speed of light.  Expressing this as a fraction of the Eddington luminosity of the black hole yields

\begin{equation}
\frac{L}{L_{\rmn Edd}} = \frac{ \epsilon \sigma_{\rmn T} G m n c}{2 (1 - \epsilon)c_{\rmn s}^3}\mbox{\ ,}
\end{equation} 
where $\sigma_{\rmn T}$ is the Thomson cross section.  With equation (20), we thus find

\begin{equation}
\frac{L}{L_{\rmn Edd}} = \frac{ \epsilon \sigma_{\rmn T} G m_{\rmn H}^{\frac{3}{2}} n c}{2 (1 - \epsilon)(3 k_{\rmn B} T)^{\frac{3}{2}}} \left(\frac{1}{m_{\rmn i}} - 
   \frac{2 \pi (1-\epsilon)m_{\rmn H}^{\frac{5}{2}}   G^2  n t}{(3 k_{\rmn B} T)^{\frac{3}{2}}   }\right)^{-1}  \mbox{\ .}
\end{equation} 

\begin{figure}
\vspace{2pt}
\epsfig{file=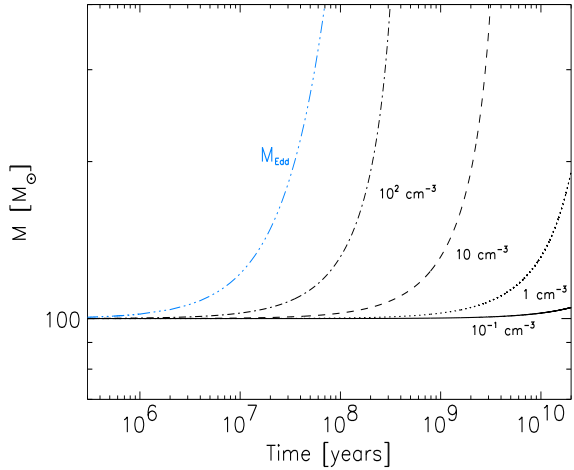,width=8.5cm,height=7.cm}
\caption{
The mass of an initially 100 ${\rmn M}_{\odot}$ black hole as a function of time, assuming the black hole accretes gas at a temperature of 200 K and at a constant 
density.  The solid, dotted, dashed, and dotted-dashed lines show cases with densities of 0.1 cm$^{-3}$, 1 cm$^{-3}$, 10 cm$^{-3}$, and 100 
cm$^{-3}$, respectively.  The triple dot-dashed line shows the mass of the black hole as a function of time, assuming that the black hole accretes at the 
Eddington limit.  
Clearly, for the black hole to begin accreting at near the Eddington limit while its mass is still of the order of 100 ${\rmn M}_{\odot}$, it must accrete gas that has a 
density $> 10^2$~cm$^{-3}$. }
\end{figure}

\begin{figure}
\vspace{2pt}
\epsfig{file=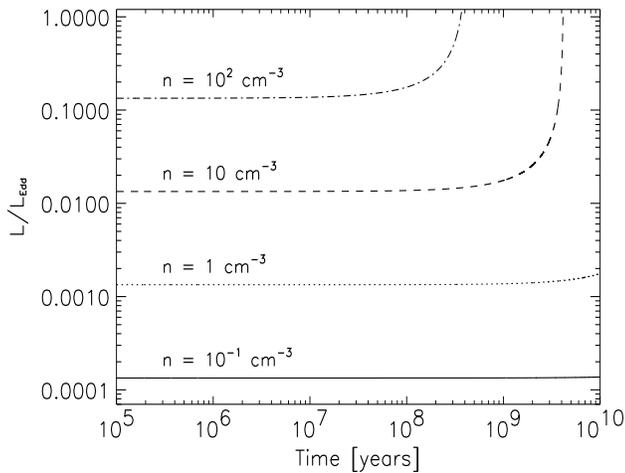,width=8.5cm,height=7.cm}
\caption{
The luminosity emitted by a miniquasar fueled by accretion onto an initially 100 ${\rmn M}_{\odot}$ black hole, expressed as a fraction of the Eddington luminosity 
of the black hole, as a function of time.  The densities assumed for the accreting gas are the same as those of Fig.~9, and the temperature is again assumed to be 
200 K. 
}
\end{figure}

To use our analytical argument to estimate the early growth rate
of the first black holes, we need to determine the densities and
temperatures of the accreting gas. We can derive their typical range of
values from our simulations of the thermal and dynamical evolution of the
gas in the vicinity of the newly formed MBH (in Sections 3 and 4).
We take it here 
that the accretion of DM particles contributes a negligible amount to the overall mass accretion rate, making the assumption that DM particles are collisionless.
Consequently, DM particles within the virialized halo in which the black hole resides cannot lose angular momentum efficiently enough to fall into the MBH.  If, 
however, DM particles were self-interacting and were thus not collisionless, the accretion of dark matter could enhance the growth of the black hole (see Hu et al. 2006).  

We do not consider the effects of radiative feedback on the accretion rate of the black hole, although it has been shown that at high accretion rates the radiation 
emitted from quasars can inhibit further gas infall (see Di Matteo, Springel \& Hernquist 2005; Springel, Di Matteo \& Hernquist 2005).  
Indeed, radiative feedback from the accretion of matter onto the black hole may heat the 
accreting gas which surrounds the black hole to higher temperatures, which will in turn lower the accretion rate, as can be seen from equation (19).
This assumption is partially 
justified by the low accretion rates and associated luminosities encountered in our simulations, 
although it is dependent on the uncertain efficiency with which the accreting gas may absorb the emitted radiation.  We have also assumed that the black hole is not 
moving with respect to the accreting gas, as might be the case if the black hole receives a kick at its birth.  These last two assumptions make our results for the 
accretion rates onto the black hole upper limits.         

Finally, as for our assumption that the fraction of accreted mass that is converted to radiation, $\epsilon$ = 0.1 will be an overestimate for accretion rates 
that are much below the Eddingtion accretion rate.  In general, $\epsilon$ may be much lower, and so our results for the luminosity generated by the accretion 
onto black holes should also be considered upper limits.   

Fig.~9 shows the mass of an initially 100 ${\rmn M}_{\odot}$ black hole as a function of time, using equation (20), for various densities of the accreting gas and
assuming a temperature for the accreting gas of 
200 K, which is the temperature of the highest density gas we find in our simulations of a relic H~II region merging with a pre-collapse un-ionized minihalo.  
Also shown in Fig.~9 is the mass of the
 black hole as a function of time, assuming that it accretes at the Eddington limit at all times.  
This figure clearly shows that in order for a black hole with a mass of the order of $\sim$ 100 ${\rmn M}_{\odot}$ to accrete at near the Eddington rate, it must accrete 
gas with a density $\ga$ 10$^2$ cm$^{-3}$.  Since these densities are only found inside collapsing haloes, and are much higher than those of the gas in the expanding 
relic H~II region around the black hole, for such a black hole to begin accreting at the Eddington rate it must merge with a minihalo that has a central density $\ga$ 
10$^2$ cm$^{-3}$.

In the case of accretion at the Eddington limit, the black hole attains a mass of $\sim$ 10$^9$ 
${\rmn M}_{\odot}$ in $\sim$ 800 Myr, which means that such a black hole could marginally grow to become a SMBH by redshift $z$ $\sim$ 6 and power the observed 
quasars at that redshift if it forms from a collapsed Pop~III star at a redshift of $z$ $\ga$ 30 and accretes at or near the Eddington limit 
continuously. However, for densities lower than $\sim$ 600 cm$^{-3}$, or gas temperatures higher than $\sim$ 200 K, an initially $\sim$ 100 ${\rmn M}_{\odot}$ black 
hole likely cannot  
grow fast enough to explain these observations, even if it attains mass also through mergers with other massive black holes (see Malbon 2006).  

Black holes of initially higher masses could accrete at higher rates, since the accretion rate varies as $m^2$, as 
shown in equation (19).  An initially 500 ${\rmn M}_{\odot}$ black hole would require accreting gas densities $\ga$ 10$^2$ cm$^{-3}$ at temperatures $\la$ 200 K in 
order to accrete at the Eddington limit.  Accreting at this limiting rate continuously, such a black hole could achieve a mass of 10$^9$ ${\rmn M}_{\odot}$ within 
700 Myr.  Thus, such a black hole could be the progenitor of the SMBHs at redshift $z$ $\sim$ 6, if it merges with a dense minihalo and so begins accreting at or near 
the Eddington limit starting shortly after its formation by the collapse of a Pop~III star at a redshift $z$ $\ga$ 20.   

As Fig.~8 shows, mergers of sufficiently dense and sufficiently nearby pre-collapse minihaloes with a minihalo hosting a Pop~III remnant black hole may have led to 
accretion of gas with densities $\ga$ 10$^2$ cm$^{-3}$ at temperatures of $\sim$ 200 K.  Now, as Fig.~9 shows, if a 100 ${\rmn M}_{\odot}$ remnant black hole 
accreted such high density gas during these mergers, it could have accreted the gas efficiently, the accretion rate even approaching the Eddington limit.
However, these high gas densities are confined to only the regions very near the centers of the merging minihaloes, as can be seen in Fig.~5.  Indeed, 
for the case of the merger shown in this figure, in which the peak density at the center of the merging minihalo becomes $n$ $\sim$ 10$^3$ cm$^{-3}$ after 100 Myr, 
the remnant black hole accretes only much lower density gas, residing in a region with a gas density of only $\sim$ 10$^{-2.5}$ cm$^{-3}$ at this same time in the 
course of the merger.  The accretion history of a Pop~III remnant black hole thus sensitively depends on the location of that black hole, 
as well as on the highest densities of the gas involved in the merger.  

One generic requirement for a Pop~III remnant black hole to accrete at the Eddington limit early enough to achieve a mass of $\sim$ 10$^9$ ${\rmn M}_{\odot}$ 
by $z$ $\sim$ 6 appears to be that it accrete the high density gas within a merging pre-collapse minihalo which can lie only a few hundred parsecs from the 
black hole at the time of its formation, as Fig.~8 shows.  Another requirement is that the peak density in this 
pre-collapse halo must be at least $n$ $\sim$ 10 cm$^{-3}$ at the time when the original Pop~III star stops 
emitting LW photons and collapses to become a black hole, so that the density in this halo can rise to $\ga$ 10$^2$ cm$^{-3}$.  As Fig.~9 shows, this is the density 
that the accreted gas must have in order for accretion to take place at the Eddington limit for a black hole with a mass of the order of 100 ${\rmn M}_{\odot}$.  
If either of these requirements is not satisfied, and the black hole does not merge with a sufficiently dense or nearby minihalo, then the black hole will not begin 
accreting 
at the Eddington rate before it is incorporated into a larger halo that can gravitationally bind the hot gas in the relic H~II region, and so begin to raise the 
density of this gas again.  The time scale for this to occur is $\sim$ 100 Myr (see Yoshida 2006), which means that there would be a substantial delay of the order of
the Hubble time at a redshift of $z$ $\sim$ 30 before a Pop~III relic black hole could begin efficiently accreting, unless it merges with a dense minihalo soon 
after its formation.

Using equation (23), we find time-dependent luminosities of a miniquasar fueled by a black hole remnant of a Pop~III star, expressed as a fraction of its 
Eddington luminosity, as shown in Fig.~10.  Here we have assumed the same range of values for the density, and the same temperature, of the accreted gas as 
was assumed in Fig.~9.  The same requirements for the black hole to begin accreting at the Eddington limit apply to the miniquasar, fueled by accretion onto 
the black hole, to radiate at the Eddington luminosity.  Thus, the Eddington luminosity will only be achieved within $\sim$ 100 Myr for the case of a merger 
with a neighboring pre-collapse halo with a central density above $\sim$ 10 cm$^{-3}$, and which lies within a few hundred parsecs from the black hole, 
at the time of the formation of the black hole. 
 As Fig.~10 shows, black holes that do not accrete cold gas at densities $\ga$ 10$^2$ cm$^{-3}$ will not radiate efficiently for well over a Hubble time at 
redshifts $z$ $\ga$ 20.  Therefore, it may be 
difficult for miniquasars to be strong ionizing sources which contribute substantially to reionization at redshifts $\ga$ 15, unless they are fueled by the 
accretion of dense gas onto Pop~III seed black holes that merge with dense minihaloes soon after their formation (see Madau et al. 2004; 
Ricotti \& Ostriker 2004; Kuhlen \& Madau 2005).  
We note, however, that the number density of accreting black holes at these high redshifts is unknown, and if this number density is high enough then even inefficiently radiating miniquasars may be able to substantially reionize the Universe.     

\section {Summary and Conclusions}
We have carried out three-dimensional numerical simulations of the evolution of the first relic H~II regions and of the mergers of pre-collapse minihaloes 
with the minihaloes that host the first Pop~III remnant black holes at the centers of the first relic H~II regions.  Although the first Pop~III stars which 
form in minihaloes emit enough ionizing radiation to evacuate the primordial gas that resides in these haloes and to disperse it into the IGM, we find 
that the radiative feedback from the first stars, nonetheless, does have important effects that could help to foster continued star formation inside the 
relic H~II regions that they leave behind.  Namely, the recombination of these relic H~II regions, due to their dynamical expansion, does not proceed 
to completion for over a Hubble time at the redshifts at which the first stars likely formed, $z$ $\ga$ 20.  The residual electron fraction that 
persists for long after the death of the central star in these regions allows for continued molecule formation, and thus high fractions of both 
H$_2$ and HD are produced.  

Due to the large size of the first H~II regions, of the order of a few kpc (e.g. Alvarez et al. 2006), and to the high H$_2$ 
fraction that is generated in the recombination of the ionized primordial gas after the death of the central star, of the order of 10$^{-3}$, we find that 
shielding of molecules inside the first relic H~II regions could have been significant, providing an optical depth to LW photons of the order of 10 
(see also Haiman et al. 2000; Ricotti, Gnedin \& Shull 2001; Oh \& Haiman 2002).  With the effects of continued H$_2$ formation and of high optical depth to LW 
photons taken together, we have shown that molecules inside the first relic H~II regions could have persisted at least down to redshifts of $z$ $\sim$ 15 
before the background LW flux in the Universe grew strong enough to efficiently photodissociate H$_2$ in these regions.  We have further shown that 
minihaloes with central densities of $\ga$ 10$^2$ cm$^{-3}$ that suffer complete destruction of H$_2$ molecules due to photodissociating flux from a Pop~III 
star can reform H$_2$ and collapse within 60 Myr after the death of the Pop~III star (see also Susa \& Umemura 2006; Mesinger et al. 2006).  
Thus, second-generation Pop~III stars could have formed inside relic 
H~II regions within minihaloes with central densities of at least this order of magnitude, and, furthermore, these sites of second-generation star formation 
could have been shielded from the background molecule-dissociating LW flux established by star formation elsewhere in the Universe.

In addition, we have shown that a very high fraction of HD molecules forms inside the first relic H~II regions, reaching levels of the order of 10$^{-7}$.  
As this fraction 
is well above the critical fraction of HD needed for efficient radiative cooling of the primordial gas to the temperature of the CMB, when this gas is in LTE, 
in principle, the first relic H~II regions produce sufficient HD to allow the formation of Pop~II.5 stars. However, we emphasize that the primordial gas 
in the first relic H~II regions must become gravitationally bound inside larger mass haloes in order for it to become dense enough to collapse and form stars, 
if indeed Pop~II.5 stars are to form inside these regions.  Nagakura \& Omukai (2005) investigate this possibility and find that stars could have formed from 
gas cooled by HD to temperatures $\la$ 100 K inside haloes massive enough to bind dynamically expanding relic H~II region gas.  
Three-dimensional numerical 
simulations of such larger mass systems are necessary to more reliably discern whether Pop~II.5 star formation can occur in the relic H~II regions 
surrounding the first stars.  
Interestingly, recent three-dimensional radiation-hydrodynamics simulations reported by Yoshida (2006) suggest that the fragmentation scale of HD-enriched gas 
inside relic H~II regions can become of the order of 10 ${\rmn M}_{\odot}$, which corresponds to the characteristic mass expected for Pop~II.5 stars 
(Johnson \& Bromm 2006).

The origin of the first SMBHs, which are observed to have masses of the order of 10$^9$ ${\rmn M}_{\odot}$ at $z$ $\sim$ 6, is an important and  
challenging question to address. In order for Pop~III remnant stellar black holes, with initial masses of the order of 100 ${\rmn M}_{\odot}$ 
to grow to become the observed SMBHs, they must begin accreting at or near the Eddington rate very soon after their formation.
Indeed, in order to explain the growth of supermassive black holes, much recent work draws on the assumption that this is the case (e.g. Malbon et al. 2006; 
Li et al. 2006).  In particluar, 
100 and 500 ${\rmn M}_{\odot}$ black holes must accrete at the Eddington limit for 800 and 700 Myr, respectively, in order to grow to 10$^9$ ${\rmn M}_{\odot}$.  
Additionally, recent work by Malbon et al. (2006) shows that it is indeed the accretion of gas that largely fuels the growth of massive black holes at redshifts 
$z$ $\ga$ 2,
 although at lower redshifts black hole mergers can become the dominant process for black hole growth.  
Because the radiation from the progenitor Pop~III stars which collapse to form such black holes evacuates the gas from their host haloes, mergers with neighboring
minihaloes must occur in order for the remnant black holes to accrete high enough density gas to accrete at or near the Eddington limit. 

Our results indicate that there may have been a substantial time delay 
between MBH formation and the onset of efficient accretion in situations where Pop~III seed black holes do not merge with neighboring neutral minihaloes 
with densities that are $\ga$ 10 cm$^{-3}$ at the time of black 
hole formation.  In this case, efficient accretion cannot begin until the black hole is incorporated into a halo sufficiently massive to gravitationally bind 
the hot relic H~II region gas (see Yoshida 2006), and this can take of order the Hubble time at
 $z\sim 30$, or $\sim$ 100 Myr.
It remains therefore an open question whether the SMBHs that power the $z\ga 6$ quasars could have grown from stellar remnants, or whether more
massive seed black holes were required (e.g. Bromm \& Loeb 2003; Lodato \& Natarajan 2006).
More realistic, fully cosmological simulations which accurately track the dynamics and growth of  
accreting black holes over $\sim 10^9$~yr will be required to further clarify the possible connection to the SMBHs
observed at lower redshifts. 

If the conditions we have found for accretion at the Eddington limit are not met, then the miniquasars fueled by accretion onto the Pop~III remnant black holes will 
also not emit radiation at the Eddington limit until well after the formation of the black holes.  This possibility of inefficient accretion onto these 
black holes has important consequences for theories of the reionization of the Universe by miniquasars, as it is often assumed that accretion at the 
Eddington rate is achievable for miniquasars at redshifts $z$ $\ga$ 15 (see Madau et al. 2004; Ricotti \& Ostriker 2004; Kuhlen \& Madau 2005).
Although, a sufficiently high number density of accreting black holes could still lead to significant reionization by miniquasars. 
If accretion onto miniquasars is inefficient, as discussed in Madau \& Silk (2005), the generation of the observed near-infrared background excess 
by Pop~III stars (e.g. Santos, Bromm \& Kamionkowski 2002; Kashlinsky 2005) may not necessarily imply an overproduction of X-rays emitted by miniquasars fueld 
by accretion onto Pop~III remnant black 
holes.  However, copious X-ray production could result from accretion of gas from a binary companion (see Belczynski, Bulik \& Rudak 2004; Saigo, Matsumoto \& 
Umemura 2004), 
even if the binary system resides in a low density environment.  
Future work will have to include more detailed, fully cosmological simulations 
with a prescription for radiative feedback from Pop~III remnant black holes to self-consistently address these issues.

\section*{Acknowledgments}
We would like to thank Gregory Shields and Marcelo Alvarez for helpful disscusion, and Massimo Ricotti for comments that improved the presentation of this work.  
The simulations used in this work were carried out at the Texas Advanced 
Computing Center (TACC). V.B. acknowledges support from NASA {\it Swift} grant
NNG05GH54G.


\end{document}